\documentclass[11pt,a4paper]{article}
\usepackage[hyperref]{acl2020}
\usepackage{times}
\usepackage{latexsym}
\usepackage{CJK}
\usepackage{adjustbox}
\usepackage{newfloat}
\usepackage{caption}
\usepackage{tikz}

\definecolor{Apricot}{RGB}{251,206,177}

\usepackage{boxedminipage}

\usepackage{microtype}

\aclfinalcopy 

\newcommand{\footremember}[2]{%
   \thanks{#2}
    \newcounter{#1}
    \setcounter{#1}{\value{footnote}}%
}
\newcommand{\footrecall}[1]{%
    \footnotemark[\value{#1}]%
}

\title{BERT-XML: Large Scale Automated ICD Coding Using BERT Pretraining}

\author{Zachariah Zhang\footremember{note1}{Equal contribution} \\
  NYU Langone Health \\
  \small\texttt{zz1409@nyu.edu} \\\And
  Jingshu Liu\footrecall{note1} \\
  NYU Langone Health \\
  \small\texttt{jingshu.liu@nyu.edu}\\\And
  Narges Razavian \\
  NYU Langone Health \\
  \small\texttt{narges.razavian@nyumc.org}\\}

\date{}

\begin{document}
\maketitle
\begin{abstract}
 Clinical interactions are initially recorded and documented in free text medical notes. ICD coding is the task of classifying and coding all diagnoses, symptoms and procedures associated with a patient's visit. The process is often manual and extremely time-consuming and expensive for hospitals. In this paper, we propose a machine learning model, BERT-XML, for large scale automated ICD coding from EHR notes, utilizing recently developed unsupervised pretraining that have achieved state of the art performance on a variety of NLP tasks. We train a BERT model from scratch on EHR notes, learning with vocabulary better suited for EHR tasks and  thus outperform off-the-shelf models. We adapt the BERT architecture for ICD coding with multi-label attention. While other works focus on small public medical datasets, we have produced the first large scale ICD-10 classification model using millions of EHR notes to predict thousands of unique ICD codes. 
\end{abstract}

\section{Introduction}

Information embedded in Electronic Health Records (EHR) have been a focus of healthcare community in recent years. Research aiming to provide more accurate diagnose, 
reduce patients' risk, as well as improve clinical operation efficiency have well-exploited structured EHR data, which includes demographics, disease diagnosis, procedures, medications and lab records. However, a number of studies show that information on patient health status primarily resides in the free-text clinical notes, and it is challenging to convert clinical notes fully and accurately to structured data \cite{ashfaq2019medication, guide2013capturing,cowie2017electronic}. 

Extensive prior efforts have been made on extracting and utilizing information from unstructured EHR data via traditional linguistics based methods \cite{savova2010mayo, soysal2017clamp, aronson2010overview, wu2018semehr}. With rapid developments in deep learning methods and their applications in Natural Language Processing (NLP), recent studies adopt those models to process EHR notes for supervised tasks such as disease diagnose and/or ICD\footnote{ICD, or International Statistical Classification of Diseases and Related Health Problems, is the system of classifying all diagnoses, symptoms and procedures for a patient's visit. For example, I50.3 is the code for Diastolic (congestive) heart failure. These codes need to be assigned manually by medical coders at each hospital. The process can be very expensive and time consuming, and becomes a natural target for automation.} coding \cite{flicoteaux2018ecstra, xie2018neural}. 
 Yet to the best of our knowledge, application of recently developed and vastly-successful self-supervised learning models in this domain have remained limited to very small cohorts \cite{alsentzer2019publicly},\cite{huang2019clinicalbert} and/or with non-clinical datasets \cite{lee2019biobert}. In addition, these models are adapted directly the original BERT models released in \cite{devlin2018bert} which use a vocabulary derived from a corpus of language not specific to EHR. 
 


In this work we propose BERT-XML as an effective approach to diagnose patients and extract relevant disease documentation from the free-text clinical notes. BERT (Bidirectional Encoder Representations from Transformers) \cite{devlin2018bert} utilizes unsupervised pretraining procedures to produce meaningful representation of the input sequences, and provides state of the art results across many important NLP tasks. BERT-XML combines BERT pretraining with multi-label attention \cite{you2018attentionxml}, and outperforms other baselines without self-supervised pretraining by a large margin. Additionally, the attention layer provides a natural mechanism to identify part of the text that impacts final prediction. 

Compare to other works using BERT for disease identification, we emphasize on the following aspects: 1) {\bf Large cohort pretraining. }We train BERT model from scratch on over 5 million EHR notes, and find it outperforms off-the-shelf or fine-tuned BERT using off-the-shelf vocabulary. 2) {\bf Long input sequence. }We model input sequence up to 1,024 tokens in both pre-training and prediction tasks to accommodate common EHR note size. This shows superior performance by considering information over longer span of text. 3) {\bf EHR Specific Vocabulary. } While other implementations use the vocabulary from the original BERT, we train with a vocabulary specific to EHR to build better representations of EHR notes. 4) {\bf Extreme large number of classes. }We use the 2,292 most frequent ICD-10 codes from our modeling cohort as the disease targets, and shows the model is  highly predictive of the majority of classes. This extends previous effort on disease diagnose or coding that only predict a small number of classes. 5) {\bf Novel multi-label embedding initialization. }We apply an innovative initialization method as described in \ref{BERT Multi-Label}, that greatly improves training stability of the multi-label attention.

\section{ Related Works}

\subsection{CNN, LSTM based Approaches and Attention Mechanisms}
Extensive work has been done on applying machine learning approaches to automatic ICD coding. Many of these approaches rely on variants of Convolutional Neural Networks (CNNs) and Long Short-Term Memory Networks (LSTMs). In \cite{flicoteaux2018ecstra}, authors use a text CNN as well as lexical matching to improve performance for rare ICD labels. In \cite{xu2018multimodal}, authors use an ensemble of a character level CNN, Bi-LSTM, and word level CNN to make predictions of ICD codes. Another study \cite{xie2018neural} proposes a tree-of-sequences LSTM architecture to simultaneously capture the hierarchical relationship among codes and the semantics of each code. 

Many works further incorporates the attention mechanisms as introduced in \cite{bahdanau2014neural}, to better utilize information buried in longer input sequence. In \cite{baumel2018multi}, the authors introduce a Hierarchical Attention bidirectional Gated Recurrent Unit(HA-GRU) architecture. \cite{shi2017towards} uses a hierarchical combination of LSTM's to encode EHR text and then use attention with encodings of the text descriptions of ICD codes to make predictions. 

While these models have achieved impressive results, they usually fall short in modeling the complexity of EHR data in terms of the number of ICD codes predicted. For example, \cite{shi2017towards} limited their predictions to the 50 most frequent codes and \cite{xu2018multimodal} predicted 32. In addition, these works do not utilize any pretraining and performance can be limited by size of labeled training samples

\subsection{Transformer Modules}

Unsupervised methods of learning word representations has been well established within the NLP community. Word2vec\cite{mikolov2013distributed} and GloVe\cite{pennington2014glove} learn vector representations of tokens from large unsupervised corpora in order to encode semantic similarities in words. However, these approaches fail to incorporate wider context into account, in learning representations of words. 

Recently, there have been several approaches developed to learn unsupervised encoders that produce contextualized word embedding such as Elmo\cite{peters2018deep} and BERT (Bidirectional Encoder Representations from Transformers) \cite{devlin2018bert}. 
 These models utilize unsupervised pretraining procedures to produce representations that can transfer well to many tasks. BERT uses self-attention modules rather than LSTMs to encode text. In addition, BERT is trained on both a masked language model task as well as a next sentence prediction task. This pretraining procedure has provided state of the art results across many important NLP tasks.

 

Inspired by the success in other domains, several works have utilized BERT models for medical tasks. \cite{shang2019pre} uses a BERT style model for medicine recommendation by learning embeddings for ICD codes. \cite{sanger2019classifying} uses BERT as well as BioBERT \cite{lee2019biobert} as base models for ICD code prediction. Clinical BERT \cite{alsentzer2019publicly} uses a BERT model fine-tuned on MIMIC III notes and discharge summaries and apply to downstream tasks.  

Transformer based architectures have led to a large increase in performance on clinical tasks. However, they rely on fine tuning off-the-shelf BERT models, whose vocabulary is very different from clinical text. For example, while clinical BERT \cite{alsentzer2019publicly} fine-tuned the model on the clinical notes, the authors did not consider expand the base BERT vocabulary to include more relevant clinical terms.  Cui \cite{cui2019pre} shows that pretraining with many out of vocabulary words can degrade quality of representations as the masked language model task becomes easier when predicting a chunked portion of a word. Moreover, notes processed are often capped at a relatively short length. For example, Clinical BERT uses a length of 128 and \cite{sanger2019classifying} truncates note length to 256. In addition most of these papers only train on the relatively small MIMIC-III \cite{johnson2016mimic} dataset which contains only 60k patients. These patients are also exclusively from critical care units which only represent a small subset patients in the EHR system for most hospitals. 


\section{Methods}

\subsection{Problem Definition}

We approach the ICD tagging task as a multi-label classification problem. We learn a function to map a sequence of input tokens $x = [ x_0, x_1, x_2, ... , x_N]$ to set of labels $y = [y_0, y_1, ... y_M ]$ where $y_j \in [0,1]$ and $M$ is the number of different ICD classes. Assume that we have a set of $N$ training samples $\{( x_i,y_i)\}^N_{i=0}$ representing EHR notes with associated ICD labels. 

\subsection{BERT Pre-training}

In this work, we use BERT to represent input text. BERT is an encoder composed of stacked transformer modules. The encoder module is based on the transformer blocks used in \cite{vaswani2017attention}, consisting of self-attention, normalization, and position-wise fully connected layers. Self-attention avoids the vanishing gradient and inductive bias associated with sequential models such as LSTM or GRU. The model is pretrained with both a masked language model task as well as a next sentence prediction task. In the former, tokens from the input sequence are randomly replaced with a [MASK] token and the model learns to predict the removed tokens.
In the latter, the model produce a binary classification to predict if one sentence follows another, in order to better model longer term dependencies. 

Unlike many practitioners who use BERT models that have been already pretrained on a wide corpus, we trained BERT models from scratch on EHR Notes to address the following two major issues. Firstly, healthcare data contains a specific vocabulary that is not common within a general pretraining corpus, leads to many out of vocabulary(OOV) words. BERT handles this problem with WordPiece tokenization where OOV words are chunked into sub-words contained in the vocabulary. Naively fine tuning with many OOV words may lead to a decrease in the quality of the representation learned as in the masked language model task as show by Cui \cite{cui2019pre}. Models such as Clinical BERT may learn only to complete the chunked word rather than understand the wider context. The open source BERT vocabulary contains an average 49.2 OOV words per note on our dataset compared with 0.93 OOV words from our trained-from-scratch vocabulary. Secondly, the off-the-shelf BERT models only support sequence lengths up to 512, while EHR notes can contain thousands of tokens. To accommodate the longer sequence length, we trained the BERT model with 1024 sequence length instead. We found that this longer length was able to improve performance on downstream tasks. We train both a small and large architecture model whose configurations are given in table \ref{tab:bert_config}
\begin{figure}[h]
\textbf{Masked Language Model Example}

\noindent\fbox{%
    \parbox{.5\textwidth}{%
    
        \colorbox{Apricot}{review} of systems : gen : no weight loss or \colorbox{Apricot}{gain} , good general state of health , no weakness , no fatigue , no fever , \colorbox{Apricot}{good} exercise \colorbox{Apricot}{tolerance} , \colorbox{Apricot}{able} to do usual activities \colorbox{Apricot}{.} heent : head : no headache , no dizziness , no lightheadness eyes : normal vision , no redness \colorbox{Apricot}{,} no blind spots , no floaters . ears \colorbox{Apricot}{:} \colorbox{Apricot}{no} earaches , no fullness , normal hearing , no tinnitus . nose and sinuses : no colds , no stuffiness , no discharge , no hay \colorbox{Apricot}{fever} , no nosebleeds \colorbox{Apricot}{,} no \colorbox{Apricot}{sinus} trouble . mouth and pharynx : no \colorbox{Apricot}{cavities} \colorbox{Apricot}{,} no bleeding \colorbox{Apricot}{gums} , \colorbox{Apricot}{no} \colorbox{Apricot}{sore} throat , \colorbox{Apricot}{no} hoarseness . neck : no \colorbox{Apricot}{lumps} , no goiter , no neck stiffness or pain . ln : no adenopathy cardiac : \colorbox{Apricot}{no} chest pain or \colorbox{Apricot}{discomfort} no syncope , no dyspnea on exertion , no orthopnea , no pnd \colorbox{Apricot}{,} no edema , no \colorbox{Apricot}{cyanosis} , no \colorbox{Apricot}{heart} murmur , no palpitations resp \colorbox{Apricot}{:} \colorbox{Apricot}{no} pleuritic pain , no sob , no wheezing , no stridor , \colorbox{Apricot}{no} cough , no hemoptysis , no respiratory infections , no bronchitis .

    }%
}

\caption{Example of masked language model task for BERT. Colored tokens are model predictions for [MASK] tokens}
\label{fig:bertlm_example}

\end{figure}


We show sample output from our BERT model in figure \ref{fig:bertlm_example}. Our model successfully learns the structure of medical notes as well as the relationships between many different types of symptoms and medical terms.



\subsection{BERT ICD Classification Models}

\subsubsection{BERT Multi-Label Classification}
\label{bert_basic}

The standard architecture for multi-label classification using BERT is to embed a [CLS] token along with all additional inputs, yielding contextualized representations from the encoder. Assume $ H = \{h_{cls}, h_0 , h_1 , ... h_N  \}$ is the last hidden layer corresponding to the [CLS] token and input tokens $0$ through $N$, $h_{cls}$ is then directly used to predict a binary vector of labels.

\begin{equation}
  \bf{y = \sigma( W_{out}  h_{cls} )}
\end{equation}

where $y \in R^M$ ,  $W_{out}$ are learnable parameters and $\sigma()$ is the sigmoid function.

\subsubsection{BERT-XML}
\label{BERT Multi-Label}

{\bf Multi-Label Attention}

One drawback of using the standard BERT multi-label classification approach is that the [CLS] vector of the last hidden layer has limited capacity, especially when the number of labels to classify is large. We experiment with the multi-label attention output layer from AttentionXML \cite{you2018attentionxml}, and find it improves performance on the prediction task. This module takes a sequence of contextualized word embeddings from BERT $ H = \{ h_0 , h_1 , ... h_N  \}$ as inputs. We calculate the prediction for each label $y_j$ using the attention mechanism shown below.

\begin{equation}
  \bf{a_{ij} = \frac{exp(  \langle h_i, l_j  \rangle)}{ \sum_{i=0}^{N} exp(  \langle h_i, l_j  \rangle) }}
\end{equation}
\begin{equation}
  \bf{c_j = \sum_{i=0}^{N} a_{ij} h_i }
\end{equation}

\begin{equation}
  \bf{y_j =   \sigma( W_a  relu( W_b c_j ))}
\end{equation}

Where $l_j$ is the vector of attention parameters corresponding to label $j$. $W_a$ and $W_b$ are shared between labels and are learnable parameters.

{\bf Semantic Label Embedding}

We notice randomly initialized multi-label attention takes long to start learning. Rather, we use the idea of semantic label embeddings \cite{pappas2019gile} to initialize the embeddings of each label $L = \{ l_0 , l_1 , ... l_M  \}$ with the BERT encoding of the plain text description of the associated ICD code. We take the mean of the BERT embeddings of each token in the description. We find this greatly increases stability during training. 


\subsection{Baseline Models}
\subsubsection{Logistic Regression}

A logistic regression model is trained with bag-of-words features. We evaluated L1 regularization with different penalty coefficients but did not find improvement in performance. We report the vanilla logistic regression model performance in table \ref{tab:results}. 

\subsubsection{Multi-Head Attention}
We then trained a bi-LSTM model with a multi-head attention layer as suggested in \cite{vaswani2017attention}. Assume $H = \{h_0, h_1, ..., h_n\}$ is the hidden layer corresponding to input tokens $0$ through $n$ from the bi-LSTM, concatenating the forward and backward nodes. The prediction of each label is calculated as below:
\begin{equation}
  {\bf a_{ik} = \frac{exp(  \langle h_i, q_{k}  \rangle)}{ \sum_{i=0}^{n} exp(  \langle h_i, q_{k}  \rangle) }}
\end{equation}
\begin{equation}
  {\bf c_k = (\sum_{i=0}^{n} a_{ik} h_i ) / \sqrt{d_h} } 
\end{equation}

\begin{equation}
  {\bf \quad c =} concatenate {\bf[c_0, c_1,..., c_K]}
\end{equation}

\begin{equation}
  \bf{y =   \sigma( W_a c )}
\end{equation}

$k = 0,..., K$ is the number of heads and $d_h$ is the size of the bi-LSTM hidden layer. $q_k$ is the query vector corresponding to the $k$th head and is learnable. $W_a \in R^{M \times Kd_h}$ is the learnable output layer weight matrix. Both the query vectors and the weight matrices are initialized randomly.


\subsubsection{Other EHR BERT Models}

We compare our pretrained EHR BERT model with others models that have been released for the purpose of EHR applications. We compare our BERT model against BioBERT \cite{lee2019biobert} as well as clinical BERT \cite{alsentzer2019publicly}. We compare using the BioBERT v1.1 (+ PubMed 1M) version of the BioBERT model and Bio+Discharge Summary BERT for Clinical BERT. We use the standard multi-label output layer described in section \ref{bert_basic}. We choose to compare only with  \cite{alsentzer2019publicly} and not \cite{huang2019clinicalbert} as they are trained on very similar datasets derived from MIMIC-III using the same BERT initialization.

\section{Experiments}

\subsection{Data}

We use de-identified medical notes and diagnoses in ICD-10 codes from the Anonymous Institution EHR system. We exclude notes that are erroneously generated, student generated, belongs to miscellaneous category, as well as notes that contain fewer than 50 characters as these are often not diagnosis related. We use a total of 7.5 million notes corresponding to visits from about 1 million patients. This data is then randomly split by patient into 70/10/20 train, dev, test sets. Notes are padded to or split to chunks of a maximum length of 512 or 1,024, depending on the model. For notes that are split, the highest predicted probability per ICD code across chunks is used as the note level prediction.

We restrict the ICD codes for prediction to all codes that appear more than 1,000 times in the training set, resulting in 2,292 codes in total. In the training set, each note contains 4.46 codes on average. For each note, besides the ICD codes assigned to it via encounter diagnosis codes, we also include ICD codes related to chronic conditions as classified by AHRQ \cite{friedman2006hospital,chi2011prevalence}, that the patient has prior to that encounter. Specifically, if we observe two instances of a chronic ICD code in the same patient's records, the same code would be imputed in all records since the earliest occurrence of that code.



\subsection{BERT-Based Models}
\subsubsection{BERT Pretraining}

We trained two different BERT architectures from scratch on EHR notes in the training set. Configurations for both models are provided in Table \ref{tab:bert_config}. We use the most frequent 20K words derived from the training set for both models. In addition, we extended the max positional embedding to 1024 to better model long term dependencies across long notes. 

Models are trained for 2 complete epochs with a batch size of 32 across 4 Titan 1080 GPUs and Nvidia Apex mixed precision training. We utilize the popular HuggingFace\footnote{https://github.com/huggingface/pytorch-transformers} implementation of BERT for training. Training and development data splits are the same as the ICD prediction model. Number of epochs is selected based on dev set loss. We compare the pretrained models with those released in the original BERT paper \cite{devlin2018bert} in the downstream classification task, including the off-the-shelf BERT base uncased model and that after fine-tuning on EHR data. The original BERT models only support documents up to 512 tokens in length. In order to extend these to the same 1024 length as other models, we randomly initialize positional embeddings for positions 512 to 1024.

\subsubsection{BERT ICD Classification Models}

Models are trained with Adam optimizer \cite{kingma2014adam} with weight decay and a learning rate of 2e-5. We use a warm-up proportion of .1 during which the learning rate is increased linearly from 0 to 2e-5. After which the learning rate decays to 0 linearly throughout training. We train models for 3 epochs using batch size of 32 across 4 Titan 1080 GPUs and Nvidia mixed precision training. Learning rate and number of epochs are tuned based on AUC of the dev set.

\begin{table}
\centering
\begin{tabular}{lll}
 & \multicolumn{2}{c}{ \textbf{EHR BERT models} }  \\ \hline
  & \textbf{small} & \textbf{big} \\ \hline
hidden size            & 512            & 768          \\ \hline
\# layers              & 8              & 12           \\ \hline
\# attention heads     & 8              & 12           \\ \hline
intermediate size      & 2048           & 3072         \\ \hline
activation function    & gelu           & gelu         \\ \hline
hidden dropout     & .1             & .1           \\ \hline
attention dropout  & .1             & .1           \\ \hline
max len                & 1024           & 1024        
\end{tabular}
 \caption{configurations for from scratch BERT models. Big configuration matches the base BERT configuration from original paper but has larger max positional embedding}
 \label{tab:bert_config}

\end{table}

\subsection{Baseline Models}
All baseline models use a max input length of 512 tokens. The multi-headed attention model utilizes pretrained input embeddings with the StarSpace \cite{wu2017starspace} bag-of-word approach. We use the notes in training set as input sequence and their corresponding ICD codes as labels and train embeddings of 300 dimensions. Input embeddings are fixed in prediction task because of memory limitation. Additionally, a dropout layer is applied to the embeddings with rate of 0.1. We use a 1-layer bi-LSTM encoder of 512 hidden nodes with GRU, and 200 attention heads.

The multi-headed attention model is trained with Adam optimizer with weight decay and an initial learning rate of 1e-5. We use a batch size of 8 and trained it up to 2 epochs across 4 Titan 1080 GPUs. Hyperparameters including learning rate, drop out rate and number of epochs are tuned based on AUC of the dev set.

\subsection{Results}

For each model we report macro AUC and micro AUC. 
We found that all BERT based models far outperform non-transformer based models. In addition we find that the big EHR BERT trained from scratch outperform off-the-shelf BERT models. We believe this speaks to the benefit of pretraining using a vocabulary closer to EHR notes. In addition we find that adding multi-label attention outperforms the standard classification approach given the large number of ICD codes.

\begin{figure}
    \centering
    \includegraphics[width=0.7\linewidth]{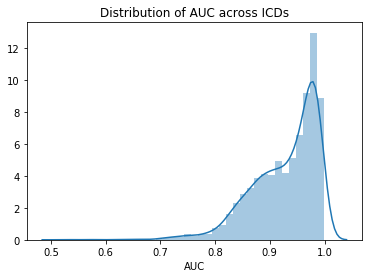}
    \caption{distributions of AUCs across ICD 10 codes}
    \label{fig:auc_dist}
\end{figure}

We analyze the performance by ICD in figure \ref{fig:auc_dist}. We achieve very high performance in many ICD classes: 467 of them have AUC of 0.98 or higher. On ICDs with low AUC value, we notice that the model can have trouble delineating closely related classes. For example, ICD G44.029-"Chronic cluster headache, not intractable" has a rather low AUC of 0.57. On closer analysis, we find that the model commonly misclassifies this ICD code with other closely related ones such as G44.329-"Chronic post-traumatic headache, not intractable". In future iterations of the model we can better adapt our output layer to the hierarchical nature of the classification problem. Detailed performance of the EHR-BERT+XML model on the test set for the top 45 frequent ICD codes is included in Table \ref{tab:results_high_freq_icds}.

Furthermore, we find that models trained with max length of 1024 outperform those of 512. EHR notes tend to be very long and this shows the value to have BERT models that can model longer length sequences for EHR applications. However, training time for the longer sequence models is roughly 3.5 times that of the shorter ones. In order to scale training and inference to longer patient histories with multiple notes it is necessary to develop faster and more memory efficient transformer models.

In addition, while the BERT based models do better than standard models on average, we see very pronounced gains in lower frequency ICDs. Table \ref{tab:results_low_freq} compares the macro AUC for all ICD codes with fewer than 2000 training examples (757 ICDs in total) of the best BERT and non-BERT models. Note that the best non-BERT model does worse on this set compare to its performance on all ICDs, while the best BERT model performs better on average on the lower frequency ones. This further illustrates the value of the unsupervised pretraining and provides good motivation to expand our analysis to even less frequent ICD codes in future work.

\begin{table*}
    \centering
        \begin{tabular}{|c|c|c| }

        \hline
    & \multicolumn{2}{c|}{AUC}   \\ \hline

     & \textbf{Micro } & \textbf{Macro }    \\ \hline
    Logistic Reg (max length 512) & 0.932 & 0.815  \\ \hline
    Multi-head Attn (max length 512) & 0.941 & 0.859  \\ \hline
    \hline
    BERT (max length 512) & 0.954 &  0.895  \\ \hline 
    BERT (max length 1024) & 0.955 & 0.898   \\ \hline 
    Finetuned BERT (max length 1024) & 0.958  & 0.903   \\ \hline
    BioBERT & 0.960 &  0.908  \\ \hline
    clinical BERT & 0.961 &  0.904  \\ \hline \hline

    EHR BERT Small (max length 512) & 0.959 &  0.897  \\ \hline
    EHR BERT Small (max length 1024) & 0.965 &  0.918  \\ \hline
    EHR BERT Small + XML (max length 1024)	&  0.968  &  0.924   \\ \hline
    \hline
    EHR BERT Big (max length 512) & 0.964  &  0.917 \\ \hline
    EHR BERT Big (max length 1024) & 0.968 & 0.925   \\ \hline
    EHR BERT Big + XML (max length 512)	& 0.967  &  0.919  \\ \hline
    EHR BERT Big + XML (max length 1024)  & \textbf{0.970} & \textbf{0.927}   \\ \hline

    \end{tabular}
        \caption{Test set model performance. The largest confidence interval calculated was only 4e-5 so all results shown are statistically significant.}

    \label{tab:results}
\end{table*}

\begin{table}[h]
    \centering
    \caption{Model Performance - Low Frequency ICDs}
    
    \begin{tabular}{|c | c |}
    
        \hline
     &  \textbf{Macro AUC}   \\ \hline
    Multi-head Att &  0.825  \\ \hline
     Big EHR BERT + XML & 0.933 \\ \hline
    
    \end{tabular}
    \label{tab:results_low_freq}
\end{table}


\begin{table*}
    \centering
    \caption{Individual ICD Performance for most frequent ICDs,  Big EHR BERT + XML. Count is the total positive examples we have observed in our test set.}
    \begin{tabular}{ |l  l  l || l  l  l || l  l  l |}
        \hline
\textbf{ICD-10}	&	\textbf{Count}	&	\textbf{AUC}	&	\textbf{ICD-10}	&	\textbf{Count}	&	\textbf{AUC}	&	\textbf{ICD-10}	&	\textbf{Count}	&	\textbf{AUC}	\\
        \hline
I10	    &	391298	&	0.877	&	M81.0	&	46528	&	0.868	&	I73.9	&	24992	&	0.898	\\
E78.5	&	291430	&	0.863	&	Z00.00	&	43136	&	0.988	&	F41.1	&	26032	&	0.842	\\
I25.10	&	131280	&	0.904	&	I48.0	&	40032	&	0.923	&	E11.65	&	22912	&	0.882	\\
E11.9	&	132150	&	0.874	&	Z51.11	&	42336	&	0.970	&	F17.200	&	21408	&	0.849	\\
K21.9	&	133422	&   0.816	&	G47.33	&	38592	&	0.846	&	Z23	    &	23296	&	0.977	\\
E55.9	&	114322	&	0.839	&	N40.0	&	34688	&	0.896	&	M17.0	&	19648	&	0.885	\\
E03.9	&	91072	&	0.840	&	J45.909	&	34496	&	0.835	&	M54.5	&	21296	&	0.973	\\
E66.9	&	80454	&	0.838	&	E66.01	&	30080	&	0.877	&	C50.912	&	21984	&	0.944	\\
E78.00	&	72740	&	0.862	&	N18.3	&	28784	&	0.888	&	M06.9	&	18160	&	0.913	\\
F41.9	&	71836	&	0.835	&	I48.2	&	26592	&	0.936	&	C50.911	&	22544	&	0.945	\\
F32.9	&	68172	&	0.824	&	Z95.0	&	24592	&	0.930	&	C50.919	&	22880	&	0.950	\\
I48.91	&	61056	&	0.922	&	G62.9	&	25632	&	0.853	&	R53.83	&	19616	&	0.968	\\
G89.29	&	49600	&	0.838	&	M17.9	&	22992	&	0.854	&	I35.0	&	17536	&	0.917	\\
J44.9	&	48224	&	0.881	&	E78.2	&	24096	&	0.876	&	Z51.12	&	20784	&	0.963	\\
M19.90	&	47968	&	0.830	&	I34.0	&	21600	&	0.900	&	J45.20	&	18848	&	0.856	\\
        \hline
    \end{tabular}
    \label{tab:results_high_freq_icds}
\end{table*}

\subsection{Visualization}

For many machine learning applications, it is important for users to be able to understand how the model comes to the predictions, especially in healthcare industry where decisions have serious implications for patients. To understand our predictions, we show the attention weights of the XML output layer for each of the classes. In figure \ref{fig:attn_vis} we show attention weights corresponding to a note coded with right hip fracture. The model successfully identify key terms such as 'right hip pain', 'hip pain' and 's/p labral'.  

\begin{figure*}
\centering
\includegraphics[width=1\linewidth]{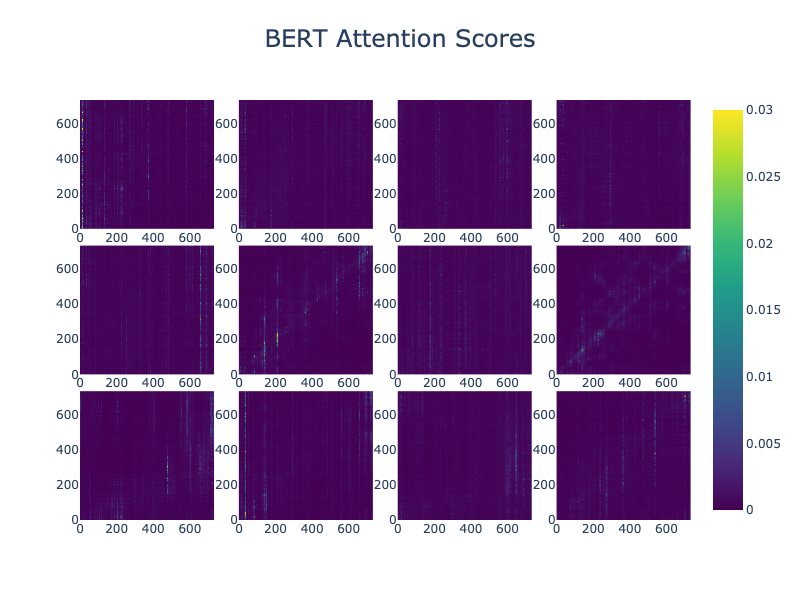}

\caption[]{ The attention weights of each head for each head in the last layer of the BERT encoder. Brighter color denotes higher attention score. We see some heads specialize in modeling local information(row 2, column 2) while some specialize in passing global information (row 1, column 1) }
\label{fig:attn_bert}
\end{figure*}

In addition, we examine the attention weights between tokens in the BERT encoder. In figure \ref{fig:attn_bert} we show the attention scores between each word of the note for the final layer of the BERT encoder of a note with 735 tokens. We observe that, while probability mass tends to concentrate between sequentially close tokens, there is a significant amount of probability mass that comes from far away tokens. In addition we see specialisation of different heads. We see that head 0 (row 1, column 1 in figure \ref{fig:attn_bert}) tends to capture long range contextual information such as the note type and encounter type which are typically listed at the beginning of each note. In addition head 5 (row 1, column 1 in figure \ref{fig:attn_bert}) models local information. We believe some of the increase in performance can be attributed to long range modeling of contextual information.

\begin{figure}
\begin{minipage}{.5\textwidth}
	
	\textbf{Prediction Visualization : Right Hip Fracture}
	
	\begin{CJK*}{UTF8}{gbsn}
{\setlength{\fboxsep}{0pt}

\colorbox{white!0}{\parbox{0.9\textwidth}{
\colorbox{red!30.44732}{\strut physical} \colorbox{red!0.0}{\strut therapy} \colorbox{red!14.068153}{\strut progress} \colorbox{red!15.188755}{\strut note} \colorbox{red!13.6379795}{\strut referring} \colorbox{red!14.236871}{\strut physician} \colorbox{red!0.0}{\strut :}  \colorbox{red!14.094721}{\strut name}  \colorbox{red!13.717354}{\strut ,}  \colorbox{red!0.0}{\strut name}  \colorbox{red!12.165296}{\strut ,} \colorbox{red!29.51839}{\strut md} \colorbox{red!15.073951}{\strut primary} \colorbox{red!0.0}{\strut care} \colorbox{red!13.679304}{\strut physician} \colorbox{red!14.657052}{\strut :}  \colorbox{red!13.962719}{\strut name}   \colorbox{red!13.61347}{\strut name}   \colorbox{red!13.2176075}{\strut name} \colorbox{red!13.458335}{\strut medical} \colorbox{red!15.3118305}{\strut diagnosis} \colorbox{red!15.205672}{\strut :} \colorbox{red!13.093678}{\strut icd} \colorbox{red!14.097309}{\strut -} \colorbox{red!13.760813}{\strut nn} \colorbox{red!12.852429}{\strut -} \colorbox{red!12.609392}{\strut cm} \colorbox{red!14.125863}{\strut icd} \colorbox{red!13.969463}{\strut -} \colorbox{red!30.894476}{\strut n} \colorbox{red!0.0}{\strut -} \colorbox{red!0.0}{\strut cm} \colorbox{red!30.347534}{\strut n} \colorbox{red!0.0}{\strut .} \colorbox{red!29.07218}{\strut right} \colorbox{red!33.19039}{\strut hip} \colorbox{red!30.484507}{\strut pain} \colorbox{red!33.37859}{\strut mnn} \colorbox{red!14.02544}{\strut .} \colorbox{red!30.492844}{\strut nnn} \colorbox{red!31.779718}{\strut nnn} \colorbox{red!14.090856}{\strut .} \colorbox{red!14.012268}{\strut nn} \colorbox{red!13.896721}{\strut treatment} \colorbox{red!14.029116}{\strut diagnosis} \colorbox{red!14.391733}{\strut :} \colorbox{red!13.56322}{\strut r} \colorbox{red!32.821236}{\strut hip} \colorbox{red!26.664654}{\strut pain} \colorbox{red!0.0}{\strut ,} \colorbox{red!32.06215}{\strut s} \colorbox{red!12.259645}{\strut /} \colorbox{red!30.894257}{\strut p} \colorbox{red!23.794245}{\strut labral} \colorbox{red!0.0}{\strut repair} \colorbox{red!15.386718}{\strut with} \colorbox{red!12.175803}{\strut [UNK]} \colorbox{red!15.085503}{\strut primary} \colorbox{red!33.434395}{\strut insurance} \colorbox{red!15.079401}{\strut :} \colorbox{red!13.960744}{\strut [UNK]} \colorbox{red!14.262858}{\strut group} \colorbox{red!13.239438}{\strut subscriber} \colorbox{red!19.415375}{\strut number} \colorbox{red!0.0}{\strut :} \colorbox{red!13.6664505}{\strut @} \colorbox{red!15.314127}{\strut subnum} \colorbox{red!13.618032}{\strut @} \colorbox{red!31.423014}{\strut secondary} \colorbox{red!32.94709}{\strut insurance} \colorbox{red!14.325375}{\strut :} \colorbox{red!32.264465}{\strut n} \colorbox{red!14.932378}{\strut /} \colorbox{red!14.499923}{\strut a} \colorbox{red!15.212954}{\strut primary} \colorbox{red!13.135007}{\strut language} \colorbox{red!14.287473}{\strut spoken} \colorbox{red!13.967423}{\strut :} \colorbox{red!15.130727}{\strut english} \colorbox{red!12.228764}{\strut [UNK]} \colorbox{red!13.860617}{\strut nn} \colorbox{red!13.090999}{\strut [UNK]} \colorbox{red!14.132845}{\strut interpreter} \colorbox{red!32.603928}{\strut present} \colorbox{red!14.568999}{\strut :} \colorbox{red!0.0}{\strut no} \colorbox{red!13.379559}{\strut any} \colorbox{red!13.816178}{\strut relevant} \colorbox{red!30.99852}{\strut changes} \colorbox{red!31.228186}{\strut to} \colorbox{red!13.646492}{\strut medical} \colorbox{red!0.0}{\strut status} \colorbox{red!14.818484}{\strut :} \colorbox{red!15.086973}{\strut no} \colorbox{red!0.0}{\strut recent} \colorbox{red!14.412281}{\strut falls} \colorbox{red!0.0}{\strut :} \colorbox{red!13.996432}{\strut no} \colorbox{red!14.006453}{\strut precautions} \colorbox{red!14.981059}{\strut :} \colorbox{red!13.499345}{\strut see} \colorbox{red!12.934329}{\strut surgical} \colorbox{red!15.255713}{\strut protocol} \colorbox{red!31.694593}{\strut in} \colorbox{red!22.146816}{\strut media} \colorbox{red!13.997249}{\strut file} \colorbox{red!13.727665}{\strut ,} \colorbox{red!13.836242}{\strut currently} \colorbox{red!29.606037}{\strut phase} \colorbox{red!13.374428}{\strut ii}  

}
}}
\end{CJK*}

\end{minipage}
\caption[]{ visualization of XML-BERT attention layer. Darker colors correspond to higher softmax value }
\label{fig:attn_vis}

\end{figure}

\section{Conclusion}

Automatic ICD coding from medical notes has high value to clinicians,  healthcare providers as well as researchers. Not only does auto-coding have high potential in cost- and time-saving, but more accurate and consistent ICD coding is necessary to facilitate patient care and improve all downstream healthcare EHR based research.

We have developed a model for ICD classification that leverages the most recent developments in NLP with BERT as well as multi-label attention. Our model achieves state of the art results using a large dataset of real EHR data across many ICDs. In addition we find that our domain specific BERT model is able to outperform open domain BERT models by modeling longer sequences as well as using a specific EHR vocabulary to overcome the WordPiece tokenizer problem. 
Our model is able to get impressive results on low frequency ICDs and we plan on expanding our model to more classes in future works. In addition, we plan on adapting our model to address the hierarchical nature of ICDs as well as developing memory efficient models that can support inference across longer sequences.

\medskip

\newpage



\bibliography{anthology,acl2020}
\bibliographystyle{acl_natbib}


Upon acceptance, the appendices come after the references, as shown here.


\end{document}